\documentclass[conference,table]{IEEEtran}

\usepackage{algorithm}
\usepackage{algorithmic}
\usepackage{amsmath,amssymb,amsfonts}
\usepackage{balance}
\usepackage{booktabs}
\usepackage{caption}
\usepackage{changes}
\usepackage{cite}
\usepackage{dirtytalk}
\usepackage{empheq}
\usepackage{enumitem}
\usepackage{environ}
\usepackage{graphicx}
\usepackage[utf8]{inputenc}
\usepackage[switch]{lineno}
\usepackage{makecell}
\usepackage{textcomp}
\usepackage{todonotes}
\usepackage[strings]{underscore}
\usepackage{url}

\newcommand{\replicationUrl}[1]{\url{https://doi.org/10.5281/zenodo.3836691}}

\definechangesauthor[name={Andreas}, color=red]{Andreas}
\definechangesauthor[name={Neil}, color=orange]{Neil}
\definechangesauthor[name={Peggy}, color=green]{Peggy}

\title{Code Duplication and Reuse in Jupyter Notebooks}
\author{
    \IEEEauthorblockN{Andreas P. Koenzen}
    \IEEEauthorblockA{\textit{Department of Computer Science}\\
    \textit{University of Victoria}\\Victoria, Canada\\akoenzen@uvic.ca}
    \and
    \IEEEauthorblockN{Neil A. Ernst}
    \IEEEauthorblockA{\textit{Department of Computer Science}\\
    \textit{University of Victoria}\\Victoria, Canada\\nernst@uvic.ca}
    \and
    \IEEEauthorblockN{Margaret-Anne D. Storey}
    \IEEEauthorblockA{\textit{Department of Computer Science}\\
    \textit{University of Victoria}\\Victoria, Canada\\mstorey@uvic.ca}
}
\IEEEoverridecommandlockouts
\IEEEpubid{\makebox[\columnwidth]{978-1-7281-6901-9/20/\$31.00 \copyright2020 IEEE \hfill} \hspace{\columnsep}\makebox[\columnwidth]{ }}

\begin{document}

\maketitle

\IEEEpubidadjcol

\begin{abstract}
Duplicating one's own code makes it faster to write software. This expediency is particularly valuable for users of computational notebooks. Duplication allows notebook users to quickly test hypotheses and iterate over data. In this paper, we explore how much, how and from where code duplication occurs in computational notebooks, and identify potential barriers to code reuse. Previous work in the area of computational notebooks describes developers' motivations for reuse and duplication but does not show how much reuse occurs or which barriers they face when reusing code. To address this gap, we first analyzed GitHub repositories for code duplicates contained in a repository's Jupyter notebooks, and then conducted an observational user study of code reuse, where participants solved specific tasks using notebooks. Our findings reveal that repositories in our sample have a mean self-duplication rate of 7.6\%. However, in our user study, few participants duplicated their own code, preferring to reuse code from online sources.
\end{abstract}

%
%

%
%
\section{Introduction}
Computational notebooks have become the preferred tool for users exploring and analyzing data. Their power, versatility and ease of use have made this new medium of computation the de facto standard for data exploration \cite{Perkel:2018}. During intensive data exploration sessions, users tend to generate great numbers of artifacts \cite{Kandel:2012}. By reusing these artifacts---in the form of Jupyter code cells---users can expedite experimentation and test hypotheses faster \cite{Kery:2017:1, Brandt:2008}. Despite the fact that software engineering best practices include avoiding code duplication whenever possible \cite{Roy:2007, Fowler:2018}, it is common behaviour with Jupyter notebooks as it is especially easy to duplicate cells, make minor modifications, and then execute them \cite{Kery:2018:2, Rule:2019}.

This form of code reuse expedites data exploration but creates notebooks that are hard to read, maintain and debug. The recommended way to reuse code is to create modules, which are standalone code files (e.g., written in Python) that can be imported locally into a notebook \cite{Rule:2019}. Unfortunately, it is reported that only about $10\%$ of notebooks contain such local imports (those imported from the repository directory) \cite{Pimentel:2019}. Hence, there is a great amount of code in notebooks for which there is no provenance, and understanding where code in notebooks originates and how it is reused is important if we want to create new tools for this environment.

Previous work in the area of computational notebooks describes developers' motivations for reuse and duplication but does not show how much reuse occurs or which barriers they face when reusing code. To address this gap, we first analyzed GitHub repositories for code duplicates contained in Jupyter notebooks, and then conducted an observational user study where participants solved specific tasks using notebooks. In our first study, we focused explicitly on code duplicates. Our definition of code duplicates is that of Roy and Cordy: \say{\textit{snippets of code copied and pasted with or without modifications, intentionally reused in order save time and effort}} \cite{Roy:2007}, although there is still some debate as to what exactly a clone is \cite{Koschke:2007}.

Given the often transient nature of notebooks, combined with the fast-paced nature of data exploration, we hypothesized that code duplication happens often in Jupyter notebooks and that it might even be useful for reducing time between ideas and results while exploring data. We know from software engineering research that \say{\textit{Cloning can be a good strategy if you have the right tools in place. Let programmers copy and adjust, and then let tools factor out the differences with appropriate mechanisms.}} \cite{Koschke:2007} We argue that code duplication can be beneficial for Jupyter notebooks with the support of the \say{right tools}.

Code duplicates---also known as code clones---have been studied extensively in software engineering, and research shows that a significant number of software systems contain code clones\footnote{In this paper, we use the terms \textit{clone} and \textit{duplicate} interchangeably.} \cite{Roy:2007, Uchida:2005}. No such study exists for computational notebooks.

We differentiate between \textit{code duplication} (artifact) and \textit{code reuse} (behaviour). We analyzed code duplication inside repositories and not across them. Hence, in this paper we use the term \textit{code duplicate} to signal code that is contained and replicated in a single repository. Although notebooks support cells of multiple types (including code and markdown text), we focused our study on code cells.

Our study focused on three research questions:
\begin{itemize}[leftmargin=1.0em]
    \item[] \textbf{RQ1:} How much cell code duplication occurs in Jupyter notebooks?
    \item[] \textbf{RQ2:} How does cell code reuse happen in Jupyter notebooks?
    \item[] \textbf{RQ3:} What are the preferred sources for code reuse in Jupyter notebooks?
\end{itemize}

We conducted two studies to answer these questions. In the first study, we mined GitHub repositories containing Jupyter notebooks, looking for code duplicates and near-duplicates. We focused on a random sample of $1,000$ GitHub repositories each containing at least one Jupyter notebook. For the second study, we designed an observational lab study ($n = 8$) where we observed participants while they solved a particular set of tasks.

Our results show that approximately 1 in 13 code cells in Jupyter notebooks are duplicates, and the nature of these duplicated snippets varies between 4 main categories: visualization ($21\%$), machine learning ($15\%$), the definition of functions ($12\%$) and data science ($9\%$). Our second study shows that the preferred method of reuse is through web browsing, mostly through tutorial sites ($35\%$), API documentation ($32\%$) and Stack Overflow ($14\%$).

%
%
\section{Background and Related Work}
Computational notebooks are a relatively new interactive computational paradigm that allows users to interleave code and text via a web interface. Programming code is introduced and segmented into \textit{code cells} that are executed in a \textit{kernel} (Python, R, Julia, C++, other) with computation output/results returned to the web interface for display. This new way of computation makes sharing and coding easy for programming newcomers as users do not need to compile code or deal with low-level configurations. Several services currently offer computational notebooks: Google Colab \cite{GoogleColab:0000} \& Cloud AI Platform \cite{GoogleCloud:0000}, Azure Notebooks \cite{Azure:0000}, Databricks \cite{Databricks:0000}, nteract \cite{nteract:0000}, Apache Zeppelin \cite{Zeppelin:0000}, to name a few. These services provide even more abstraction by taking care of \textit{kernel} configurations and just providing one for the user to select and use. In this section, we will discuss code reuse within this medium of computation.

\subsection{Code Duplication and Reuse}
Code cloning or duplication is considered a bad practice or bad smell in software engineering as described by Fowler \cite{Fowler:2018} as it is believed to cause maintainability issues \cite{Juergens:2009, Roy:2007}. However, other studies that analyzed the impact and damage of code clones have provided evidence that the problem might be less severe than what was originally estimated \cite{Thummalapenta:2009}. It is always preferable, and in fact it is highly recommended as good practice, to create modules with functions that can be accessed through interface implementations. However, resorting to duplicates can sometimes simplify the development effort, especially if the goal of the code is to be used as a playground or for testing, as is the case with Jupyter notebooks \cite{Kery:2018:2}.

Previous studies in computational notebooks have analyzed how people use them, and reports show that when it comes to modularity, only about $10\%$ of Jupyter notebooks contain imports from local libraries \cite{Pimentel:2019}. This flexibility in the design of Jupyter notebooks might be due to the fact that their users are not concerned with coding best practices \cite{Wang:2019} but with ease of use. Or maybe it may be due to the fact that users of Jupyter notebooks prioritize finding a solution over writing high quality code, as reported in a study by Kery \textit{et al.} \cite{Kery:2017:3}.

Although coding best practices are not paramount for users of Jupyter notebooks, there are projects that try to shift that attitude into one more oriented towards reusability and modularity. One of these projects is Papermill \cite{Papermill:0000}, an nteract \cite{nteract:0000} library for passing parameters to Jupyter Notebooks. It lets users reuse a notebook by passing specific parameters at run-time, allowing one to try multiple approaches without needing to create extra cells. This form of reuse is particularly necessary for computational notebooks since they allow users to execute notebooks from the command line just like a regular script, and to collect computation results using different mediums (local files, S3, and others).

Another form of reuse that is widely used is the practice of adding snippets of code to notebooks with the click of a mouse. This form of reuse entails a local library of snippets, from which the user can read and write snippets. Google Colab\cite{GoogleColab:0000} offers a function for users to specify a notebook where reusable snippets of code reside and from where users can reuse with a simple click. This form of quick duplication and reuse has been defined by users of notebooks on Stack Overflow and other internet forums as a \say{\textit{super needed feature}} and as a \say{\textit{useful way to insert small, reusable code chunks into a notebook with a single click}}.

Other forms of reuse have been studied before. Kery and Myers \cite{Kery:2017:1} reported that developers relied extensively on copying versions of their files to support their data exploration sessions. Others have suggested new tools that expedite exploration by enabling better access to previous artifacts and exploration history \cite{Kery:2017:3, Kery:2018:3}. These tools have focused on internal \textit{in-notebook} code duplication and reuse, using past cells and a notebook's history as a source of reuse.

Chattopadhyay \textit{et al.} \cite{Chattopadhyay:2020} surveyed Microsoft data scientists about notebook pain points. One of the reported pain points is the difficulty of exploring and analyzing code, which results in continual copy \& paste cycles. Their participants also ranked activities based on importance, and \textit{Reuse Existing} code was labeled as at least important $94\%$ of the time.

It is also worth mentioning that reuse is not limited to any specific source. It can come from either web pages, other notebooks, or from version control system (VCS) repositories (e.g. \textit{git}, SVN and others). Other studies have investigated version control systems supporting analysts' exploration of data, like studies conducted by Kery \cite{Kery:2017:2}, where participants reported not relying on VCS for their exploration sessions despite using them often for other tasks.

As it is with VCSes, reusing code from other Jupyter notebooks presents some issues as well. Studies have reported difficulties choosing easily identifiable names for files and folders \cite{Kery:2017:3}, which generate confusion when trying to find relevant snippets of code. Imagine a data analyst creating a different notebook for each analysis path they decide to take, e.g., they may well end up with many different notebooks named \textit{hypothesis\_1.ipynb}, \textit{hypothesis\_2.ipynb} and so on \cite{Rule:2018:1, Chattopadhyay:2020}. This type of versioning presents many problems when it comes to finding useful snippets of code, including how to distinguish one exploration path from the other, and how to quickly know which one contains the snippet we are looking for. One way to solve these problems would be to provide for longer names that could describe more in depth what a notebook contains or is about, but that introduces new problems in itself, namely, longer and more convoluted names. Another solution would be to allow users to traverse previous notebooks more easily.

In light of these issues with reuse and duplication, we set out to understand \textit{how}, \textit{how much} and \textit{from where} reuse and duplication occurs within Jupyter projects and repositories.

%
%
\section{Study 1: Counting Jupyter Code Cell Duplicates on GitHub}
In order to better understand code duplication in computational notebooks, we decided to mine GitHub repositories using the data set created by Rule in \cite{Rule:2018:1}. Rule's study retrieved $1.25$ million notebooks from GitHub, which they estimate as $95\%$ of the notebooks available in 2017. We used a random sample of $1,000$ repositories provided with this data set. Jupyter notebooks are just JSON text files segmented into cells and associated metadata. Cells can be markdown (documentation and text), source code (using the kernel), images (e.g., charts), or raw data. This study focuses on source code cells---the ones with snippets of programming code.

In our first study, we set out to answer \textit{RQ1}: \textit{How much cell code duplication occurs in Jupyter notebooks?}

\subsection{Code Duplicates}
According to Roy and Cordy \cite{Roy:2007}, code duplicates can be introduced in a software system by a) copy and paste, b) by forking, and c) by design, functionality and logic reuse. They categorized duplicates into four types:
\begin{itemize}[leftmargin=1.0em]
    \item[] \textbf{Type-1:} An exact copy of a code snippet except for white spaces and comments in the source code.
    \item[] \textbf{Type-2:} A syntactically identical copy where only user-defined identifiers such as variable names are changed.
    \item[] \textbf{Type-3:} A modified copy of a Type-2 clone where statements are added, removed or modified. Also, a Type-3 clone can be viewed as a Type-2 clone with gaps in-between.
    \item[] \textbf{Type-4:} Two or more code snippets that perform the same computation but are implemented through different syntactic variants.
\end{itemize}

In this study, we focused on the first three types of duplicates. Detecting Type-4 duplicates is complex and we believed the first three types are sufficient to answer RQ1.

\subsection{Method}
We started with a random sample of $1,000$ GitHub repositories containing approximately $6,000$ notebooks from Rule \textit{et al.} \cite{Rule:2018:2}\footnote{All artifacts used for this study are provided at \replicationUrl.}. We cloned each repository and looked at the latest commit available. We then extracted all code cells from each notebook in that repository. Once we extracted all code cells from a repository, we ran our own function (Function \ref{math:duplication_ratio}) on every code cell, comparing that cell against all other cells in the repository, looking for Type-1, Type-2 and Type-3 duplicates. Based on the duplicate counts, we calculated a \textit{Repository Duplicates Ratio}, which is the ratio of duplicated code cells to total number of code cells (references to `cell' should be taken as referring to `code cells' for the remainder of the paper).

\vspace{-1.2em}
\begin{equation}
    \label{math:duplication_ratio}
    \scalebox{0.83}{
        $DR(C_1, C_2) = \dfrac{LD(C_1, C_2)}{(\log{\mathsf{avglen}(C_1, C_2)})^{\lambda_1} + (\log{\mathsf{avgloc}(C_1, C_2)} )^{\lambda_2}}$
    }
\end{equation}
\vspace{-0.6em}

Function \ref{math:duplication_ratio} returns the \textit{duplicate ratio} (DR) between two cells. It returns a real number in the interval $[0,+\infty)$. $LD(C_1, C_2)$ corresponds to the Levenshtein distance \cite{Levenshtein:1966} between cells $C_1$ and $C_2$. $\mathsf{avglen}(C_1, C_2)$ corresponds to the average \textit{number of characters} in cell $C_1$ and $C_2$, and $\mathsf{avgloc}(C_1, C_2)$ is the average \textit{number of lines of code} in cells $C_1$ and $C_2$. Parameters $\lambda_1$ and $\lambda_2$ are constants which act as weights. $\lambda_1$ weights the number of characters, and $\lambda_2$ weights cell lines of code. Setting these parameters allows us to de-emphasize short, quick print statements (few lines of code) or long blocks of text with few lines of code. We experimented with $\lambda$ settings, heuristically determining the optimal setting to be $\lambda_1 = 6$, $\lambda_2 = 8$, such that lines of code carry more weight than the number of characters.

Duplicates with a DR of $0$ are identical (Type-1 duplicates), and the bigger the DR value is, the less similar the two blocks are. We only considered code to be duplicated if it had a DR of $0.3$ or lower. We came up with that cut-off value by heuristics, experimenting with a smaller random sample, empirically assessing snippets detected as duplicates. We detected duplicates with different thresholds for the cut-off value, e.g., $0.0-0.1$, $0.1-0.2$, ..., $0.9-1.0$, and we were able to verify that at threshold $0.8-0.9$, the quality of duplicates began to decrease drastically.

We opted for a text-based/string-based method of detecting clones because it has been used effectively in other studies \cite{Duala_Ekoko:2007}. We also required cross-language support because Jupyter notebooks support multiple programming languages and kernels. The Levenshtein distance is the minimum number of operations (insertions, deletions or substitutions) required for a string to be equal to another one. This method for detecting code duplicates proved to be effective for detecting Type-1, Type-2 and Type-3 duplicates (see below), but with a highly inefficient running time of $O((n*m)!)$, where $n$ and $m$ are the lengths of $C_1$ and $C_2$ in characters. For this study, we implemented our own function (Function \ref{math:duplication_ratio}) in order to have more control in the detection of snippets. We also removed comments and leading/trailing white space from lines of code.

Finally, once we found duplicates, we randomly sampled $500$ duplicates and thematically coded the duplicated code's purpose in order to get a better understanding of the nature of duplicates. When coding the snippets, we tried to answer questions like \textit{what is its goal} and \textit{what is it trying to compute}.

\subsection{Results}
We searched for duplicates using Function \ref{math:duplication_ratio} on $897$ repositories, consisting of $6,386$ notebooks containing $78,020$ code cells. $103$ repositories were no longer available. Only $429$ contained more than $28$ code cells in total (across all notebooks in that repository). Since $28$ was the median, and the number of code cells is exponentially distributed, we discarded repositories with fewer than $28$ code cells to a) reduce the running time and b) ensure trivial repositories were not counted. From that analysis, we detected $5,872$ Type-1, Type-2 and Type-3 code duplicates in total. Our mining results show that $74\%$ ($4355$ out of $5872$) of the clones were Type-2 and Type-3, and the rest were Type-1. The number of code duplicates in a repository varies mostly between $0$ and $100$, with some outliers. We now discuss our findings for the distance between duplicates (their duplicate type), the distribution of duplicate ratio (DR), and duplicate purpose.

\subsubsection{Duplicate Type}
The Levenshtein distance (LD) between code cells follows an exponential distribution, with a median of $21$, mean of $41.08$, standard deviation of $59.66$, minimum value of $0$ and maximum value of $535$. Most duplicates detected by our algorithm were Type-1 and Type-2 (closer to zero), with a long fat-tail where some Type-3 (further away from zero) duplicates were detected.

\subsubsection{Repository Duplicates Ratio}
Duplicate ratio measures the number of duplicate code cells over the total number of code cells in a repository. It also follows an exponential distribution, with a median ratio of duplicates per repository of about $5.0\%$ with a mean and standard deviation of $\mu = 7.6\%$ (one in thirteen), $\sigma = 8.3\%$. The minimum ratio was $0\%$, i.e., a repository with no duplicates, and the maximum ratio was $47.5\%$, i.e., a repository where nearly half the code cells were duplicates.

\subsubsection{Coding of Duplicates}
\begin{figure}[ht!]
    \centering
    \captionsetup{justification=centering,margin=0cm}
    \includegraphics[width=\columnwidth]{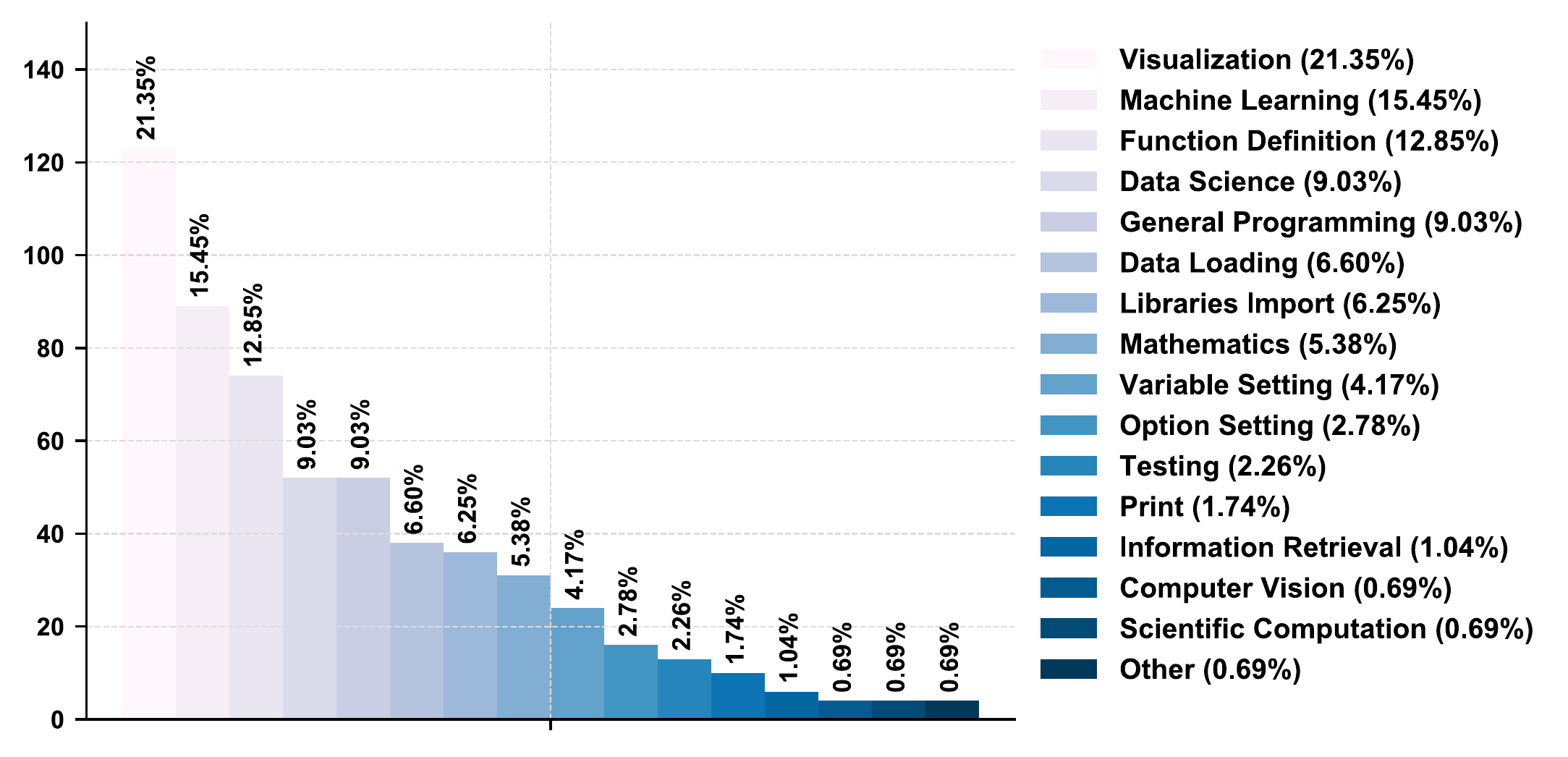}
    \caption{Inductive coding of cell code snippets marked as duplicates by our algorithm.}~\label{fig:duplicates_taxonomy}
\end{figure}

Figure \ref{fig:duplicates_taxonomy} shows the result of our inductive coding. Snippets of code that get duplicated the most within Jupyter notebooks are the ones whose main activity concerns \textit{visualization} ($21.35\%$), followed by \textit{machine learning} ($15.45\%$), \textit{definition of functions} ($12.85\%$) and \textit{data science} ($9.03\%$). We use the term \say{main activity} because some categories overlap in their activities to some degree, e.g., we merged \textit{mathematics} and \textit{statistics} together.

%
%
\section{Study 2: Observing Code Reuse Behaviour}
\begin{figure*}[ht!]
    \centering
    \captionsetup{justification=centering,margin=0cm}
    \fbox{\includegraphics[width=.98\linewidth]{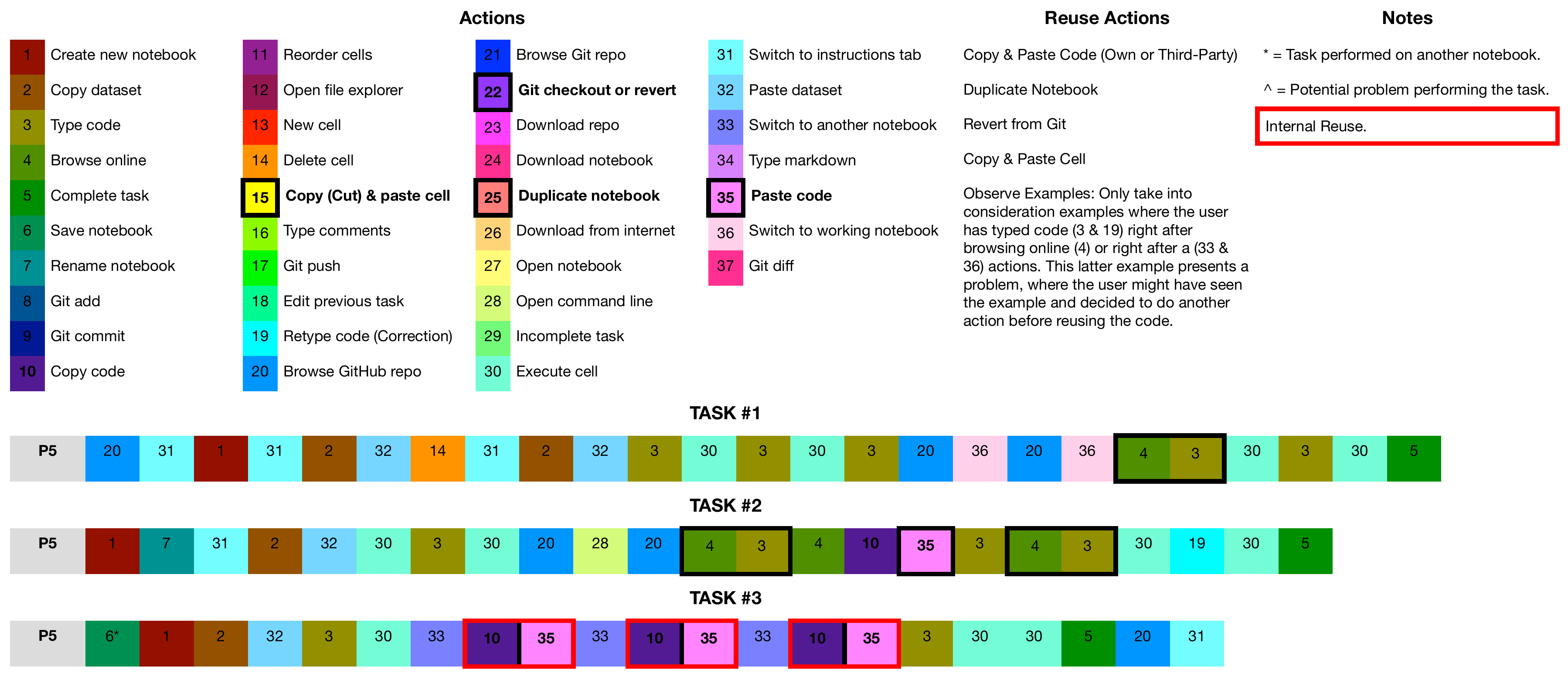}}
    \caption{Example coding of steps one of our participants (P5) made during the observational study, based on video and audio recordings.}~\label{fig:workflow}
\end{figure*}

In order to better understand how users of computational notebooks reuse code, we conducted an observational study in our lab. We observed Jupyter users reusing code using Jupyter notebooks, \textit{git} and web browsing.

We used this observational study to answer \textit{RQ2}: \textit{How does cell code reuse happen in Jupyter notebooks?} and \textit{RQ3}: \textit{What are the preferred sources for code reuse in Jupyter notebooks?}

\subsection{Method}
We observed the behaviour of eight participants (six M.Sc. and two Ph.D.). All were University students from Computer Science ($6$) and Chemistry ($2$); two were female; three last used notebooks over one month ago, five within the past day. Four reported intermediate programming skill, two advanced, and two beginner. We explicitly expressed to our participants that we were not measuring programming abilities. We recruited participants with experience with Jupyter notebooks, irrespective of their level of proficiency with programming languages. We drew this convenience sample through personal contacts and email.

Each participant was asked to solve three different tasks on our lab computers. These tasks were distributed as Jupyter notebooks according to the level of proficiency each participant reported having (Levels A, B, C, below). Each set of tasks were of varying difficulty. Each task was designed to take around $20$ minutes to complete, but participants were given more time if needed. Full instructions for how to complete each task was given in full detail on each Jupyter notebook. In total, each participant received three Jupyter notebooks with instructions for each of the three tasks\footnote{All artifacts generated for this study are provided at \replicationUrl.}. Two small data sets ($10$ elements) were also provided within each notebook with instructions. The tasks given to participants were (in the order that they were presented to the participant):
\begin{itemize}[leftmargin=1.0em]
    \item[] \textbf{Proficiency Level A}
    \begin{enumerate}
        \item Calculate the mean of a data set.
        \item Calculate the sum of all elements of a data set.
        \item Calculate the mean of data set \#1 and calculate the sum of all elements of data set \#2.
    \end{enumerate}
    \item[] \textbf{Proficiency Level B}
    \begin{enumerate}
        \item Calculate the standard deviation of a data set.
        \item Create and plot a histogram with all elements of a data set.
        \item Calculate the standard deviation of data set \#1 and create and plot a histogram with all elements of a data set.
    \end{enumerate}
    \item[] \textbf{Proficiency Level C}
    \begin{enumerate}
        \item Create a function that calculates the mean of a data set.
        \item Create a function that calculates the standard deviation of a data set.
        \item Calculate the mean of a data set using a function and write the function in the notebook. Calculate the standard deviation of a data set using a function and write the function in the notebook.
    \end{enumerate}
\end{itemize}

Tasks $1$ and $2$ were designed to be completely independent of each other, while task $3$ was an intersection of the previous two (e.g., participants could have re-used the solutions to tasks $1$ and $2$). Tasks that were based on data exploration remained fairly generic and did not rely much on external libraries.

The restrictions imposed on the participants on how to accomplish these tasks were minimal. We told them each task should be completed in a notebook different from the one given with the instructions. This allowed us to observe if users created new notebooks or reused old ones. Supported languages were Python, R and JavaScript, which the participant could choose at any time, or change in the middle of the task if needed. Participants were told they could use any resource they found online. We also linked the instructions to a GitHub repository that contained the solution to each of the three tasks in its commit history.

One researcher observed each participant during a time frame of roughly $60$ minutes. The researcher took detailed notes of the behaviours each participant displayed. Audio and screen video was also recorded. The observational study was complemented with a questionnaire and a follow-up unstructured interview. Audio, video and notes were coded for qualitative and quantitative conclusions by one researcher. Video coding was important to quantify the steps each participant took to complete each task, to understand order and to sequence and find patterns in participant behaviours. The video coding resulted in a detailed workflow analysis with tasks coded as shown in Figure \ref{fig:workflow}. Audio was transcribed and coded to derive the qualitative aspects of the answers each participant gave.

\subsection{Results}
Our participants duplicated code in a variety of ways. After analyzing the video artifacts, we created the following coding scheme to describe how they reused their code:
\begin{itemize}[leftmargin=1.0em]
    \item[] \textbf{C\&P:} Copying \& pasting lines of code.
    \item[] \textbf{CELL:} Copying \& pasting code of an entire cell (reuse from notebooks).
    \item[] \textbf{TYPE:} Typing code written in another notebook of theirs, instead of \textbf{C\&P} it.
    \item[] \textbf{DUPE:} Duplicating a notebook of their own.
    \item[] \textbf{GIT:} Reusing from \textit{git}.
    \item[] \textbf{TYPE\_ON:} A special case where participants would browse online and then type the code they extracted from the source instead of \textbf{C\&P} it.
    \item[] \textbf{NONE:} No reuse; directly enter solution from memory.
\end{itemize}

All participants reused code quite extensively, and only one participant typed from memory (NONE). Most participants reused code from online sources. Foraging for code online is a popular form of code reuse among programmers and analysts, as was pointed out by Brandt \textit{et al.} in \cite{Brandt:2009}. We also observed that participants reused code from internal sources, like other notebooks. This was expected given they had easy access to previous tasks. None of the participants decided to reuse from \textit{git} (GIT), although full solutions to each task were readily available in the local \textit{git} repository and on GitHub.

\subsubsection{Code Reuse from Other Notebooks}
Four out of eight participants decided to reuse what they did in task $1$ and $2$ for task $3$. The other four decided to perform task $3$ from scratch, even though they had the necessary code for task $3$ already implemented for task $1$ and $2$ (recall that task $3$ is deliberately structured as the union of $1$ and $2$). When asked why, one participant (P1) said: \say{\textit{Muscle memory. As a means of preserving knowledge.}}

P2 had trouble copying and pasting the code inside a cell. One explanation for this is that in Jupyter, when doing a right-click of the mouse, only copy or cut at the cell level is available. P4 and P6 both stated that they enjoyed typing.

\subsubsection{Code Reuse from External Sources}
All participants used the web extensively to assist themselves in the completion of each task. The time spent browsing online accounted on average for $18\%$ of the total time spent working on tasks (Browsing time: $\mu =$ $233$ $\pm$ $180$ seconds). We also observed that they relied on online resources like API documentation and tutorials for repetitive tasks, e.g., some participants browsed online more than once for examples on how to import the same library, even though they performed the same import just minutes before (Browsing count: $\mu =$ $9.6$ $\pm$ $3.8$ times). We noted that some participants used the web as a \textit{memory delegate} \cite{Brandt:2009}.

\bgroup
\def\arraystretch{1.3}
\begin{table}[ht!]
    \centering
    \captionsetup{justification=centering,margin=1cm}
    \begin{tabular}{cccc}
        \toprule
        \textbf{Participant} & \textbf{Time} & \textbf{Percentage} & \textbf{Count} \\
        \midrule
        \textbf{P1} & 176 sec & 30\% & 11 times \\
        \textbf{P2} & 170 sec & 21\% & 3 times \\
        \textbf{P3} & 155 sec & 13\% & 7 times \\
        \textbf{P4} & 678 sec & 31\% & 14 times \\
        \textbf{P5} & 95 sec  & 8\%  & 6 times \\
        \textbf{P6} & 178 sec & 9\%  & 13 times \\
        \textbf{P7} & 317 sec & 18\% & 14 times \\
        \textbf{P8} & 95 sec  & 9\%  & 9 times \\
        \bottomrule
    \end{tabular}
    \caption{Portion of study spent browsing online per participant.}
    \label{table:online_time}
\end{table}

Browsing and reusing common libraries in Python was also a popular resource among our participants, as all of them relied heavily on the \textit{numpy} library for solving the tasks. We noticed this reliance on external sources and libraries saved time and effort for the participants, since calculating the mean of a data set took them at most two lines of code to accomplish using the \textit{numpy} library (including the import statement) instead of using \textit{for loops}.

\begin{figure}[ht!]
    \centering
    \captionsetup{justification=centering,margin=0cm}
    \includegraphics[width=.88\columnwidth]{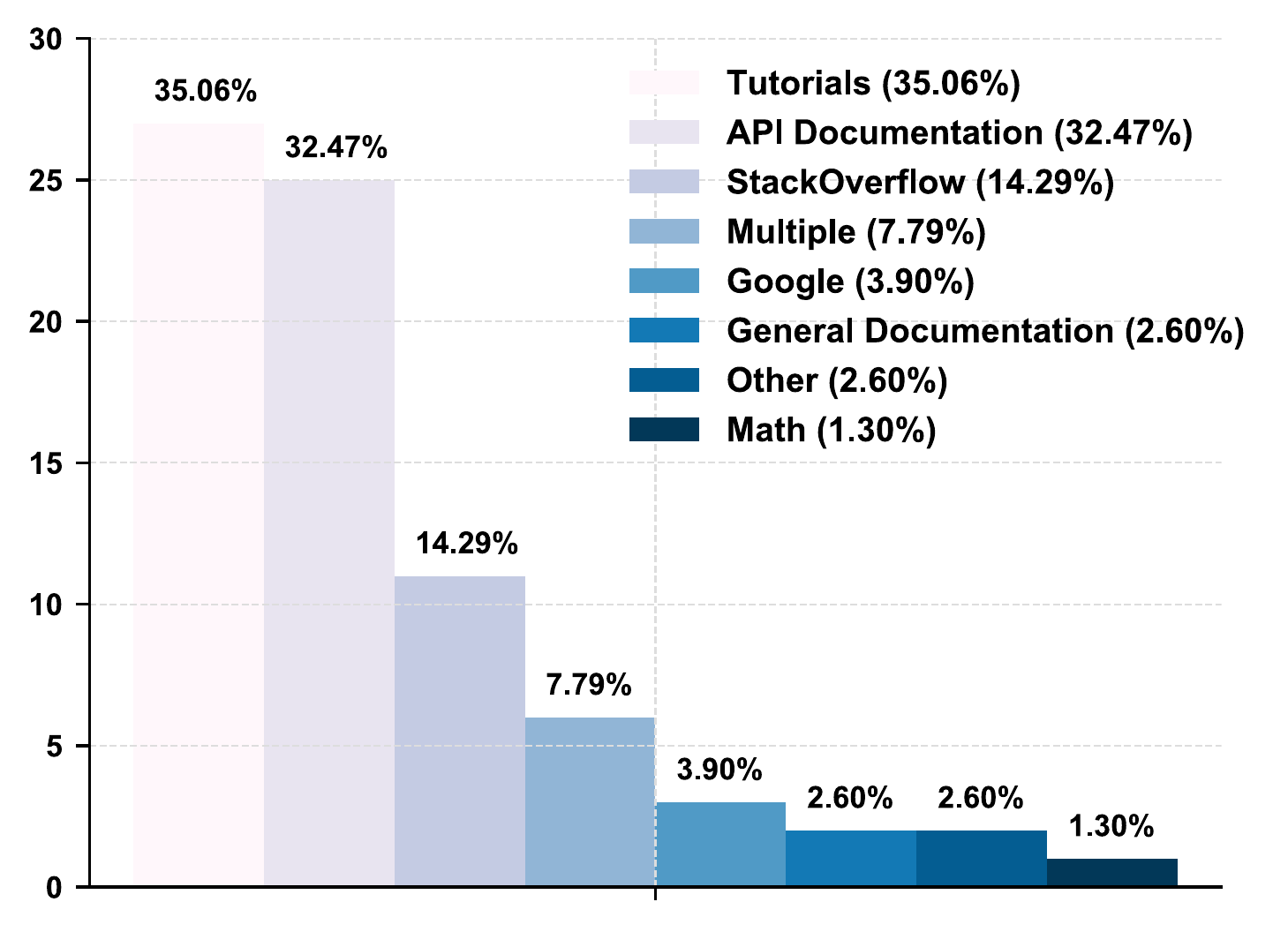}
    \caption{Inductive coding of sites participants visited while solving tasks. Note: \textit{Google} implies information taken directly from Google's results page.}~\label{fig:codes}
\end{figure}

Figure \ref{fig:codes} shows our coded responses for which sites were visited according to primary role. The two most visited sites among participants were tutorials (e.g., tutorialspoint) and API documentation, followed by Stack Overflow. Participants queries were usually something short and precise, like \say{\textit{mean numpy}} or \say{\textit{histogram numpy}}, but in some cases we observed longer, more natural language oriented queries, like \say{\textit{what is git repo? and how to use it?}} or \say{\textit{how to create Jupyter project}}.

We observed two distinct browsing habits: half the participants visited web sites nine times or fewer (Table \ref{table:online_time}), spending overall $129$ seconds on average. For example, P2 spent $21\%$ of their time browsing online, but only did this three times. They took their time skimming the web page for code reuse. The other half had $11$ or more visits, taking $337$ seconds on average. This group used web resources like an external memory aid, going back and forth between the working notebook and the online resource multiple times.

\subsubsection{Code Reuse from VCS}
We observed that although some participants went to browse for the provided solutions on the study's GitHub repository, they either did not restore them from the \textit{git} history, or lost interest after a few tries. The solutions were not readily available at the HEAD of the commit tree, so knowledge on how to traverse the commit tree and on how to move the HEAD of the tree to a particular commit or how to \textit{checkout} a particular commit was necessary in order to access the solutions. That is, above average knowledge of \textit{git} was necessary for reusing code from the local repository.

We received various answers from our participants when asked why they did not restore the provided solutions from \textit{git}. Two out of eight participants stated \say{Sufficient Knowledge} as their answer. It was implied by these participants that the tasks were not sufficiently difficult to merit restoring them from \textit{git}. And they perceived restoring from \textit{git} as far more time consuming than actually coding the task.

There was also a perception of complexity in restoring from \textit{git} as noted by one of our participants (P2). When asked why they did not restore from \textit{git}: \say{\textit{\dots if I got really stuck, then just look things up, because probably it would have taken me more time to find it in git, than actually do it myself.}}

%
%
\section{Discussion}
We used two research strategies to triangulate our understanding of code duplication and reuse. In the first study, we conducted a computational data analysis of existing GitHub notebooks. In the second study, we focused on how that duplication occurs with human observation in a lab setting. In both studies, we observed that duplicated code was important. We also identified some curious properties of version control, and finally, we revealed some patterns in external source reuse.

We look at each of these observations in turn, and conclude with how we think they influence research and practice for future work in computational notebooks.

\subsection{Code Duplicates}
Figure \ref{fig:figure5} shows an example of a Type-2 snippet of code detected by our algorithm. We can observe in the figure that both blocks of code are pretty much similar, with the exception of the variables passed as arguments to the plotting function. Before conducting our studies, we hypothesized that this type of duplicate---ones created to quickly perform some function on data---would be common in computational notebooks due to the scratch pad nature of the medium and the attitude of its users. Kery and Myers showed that users expect the notebook to be preliminary work and short-lived \cite{Kery:2018:2}.

Our computational study (Study 1) revealed that these duplicates had a median rate of $5\%$ per repository. This is an under-count, since it only looks at self-duplication, not duplication of code snippets from API usage guides or other resources. Study 2 showed that these were popular sources for reuse.

\begin{figure}[ht!]
    \centering
    \captionsetup{justification=centering,margin=0cm}
    \includegraphics[width=\columnwidth]{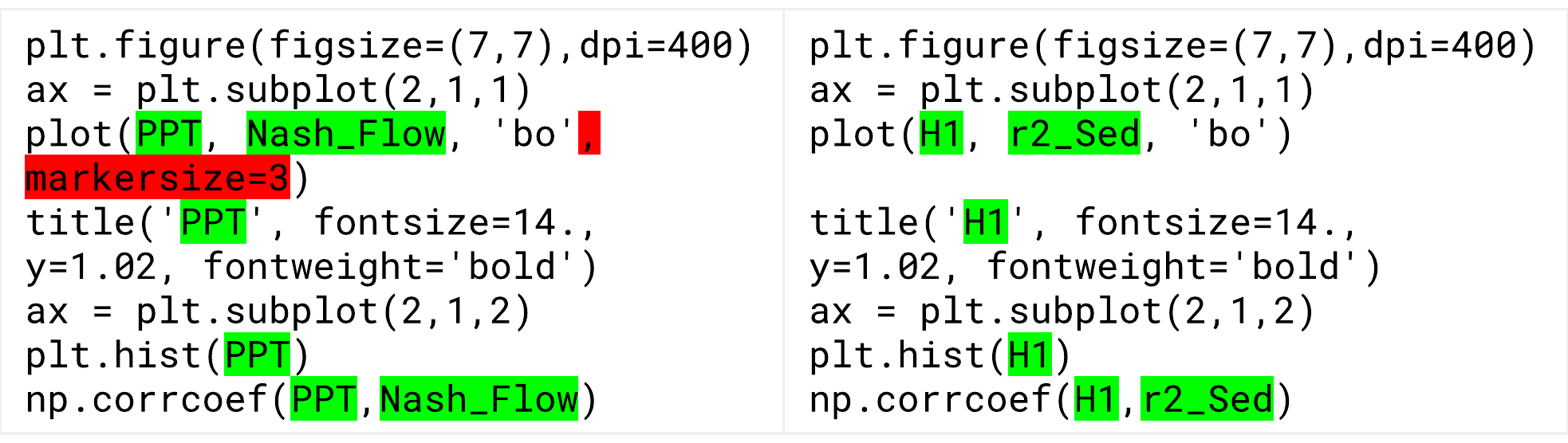}
    \caption{Example of a Type-2 duplicate detected by our algorithm with Levenshtein distance of 42 and Duplicate Ratio of 0.27. Coded as \textit{Visualization}.}~\label{fig:figure5}
\end{figure}

\textbf{Practice implications:} After conducting our study, we argue in favor of tools that support this type of reuse, whether they are offered through functions like Google Colab's \textit{Code Snippets} or similar ones designed for other types of notebooks.

\textbf{Research implications}: We observed that while this type of duplication was present in our sample, our users struggled to easily make use of previous cells, even when those cells solved exactly the same problem. This suggests that merely providing a mechanism to duplicate code has to overcome barriers of ease of use. It seems to be simpler, for some cases, to copy and paste from online sources (like Stack Overflow) than to do the same thing from one's own work (reinventing the wheel attitude).

\subsection{Use of Version Control}
Version control for notebooks has been the focus of several notebook plugins, many blog posts and feature requests (e.g., jupyterlab-git\cite{jupyterlab-git:0000}, Verdant\cite{Verdant:0000}, nbdime\cite{nbdime:0000}), and several research studies \cite{Kery:2017:3, Kery:2018:1, Chattopadhyay:2020}. Some notebook services like NextJournal\cite{NextJournal:0000} value the importance of preserving history in computational notebooks so much, that they offer automatic versioning of the notebook and related artifacts. Our first study did not examine the role of version control as a source for duplication, in part because identifying duplication from version control is tricky (since files can be renamed, sections moved, etc.). However, part of our lab study was intended to explore how version control systems (VCS), specifically \textit{git}, were used in code reuse. We saw that users struggled with the interaction model of \textit{git} and were unable to use it for duplication purposes.

\textbf{Practice implications:} Based on our limited study, version control of notebooks is less important for future code reuse than for archival purposes or collaboration. For single-user notebooks, in particular, complex tools for version control and diff might be replaced with simpler save and restore functionality like in backup interfaces (such as Apple's Time Machine \cite{Apple:0000} model).

\textbf{Research implications}: Evidence from other notebook studies \cite{Kery:2017:1, Rule:2018:2, Head:2019}, and our observations in this paper show that code duplication and reuse using \textit{version control history} is a challenge. However, \textit{git} was designed for software development on the Linux operating system, and evidence suggests the code reuse scenario is low on the list of reasons developers use version control. As Codoban \textit{et al.} reported, software developers instead use software history for debugging, program understanding, and collaboration \cite{Codoban:2015}. This use case is different than searching for previous solutions and may explain why version control tools like \textit{git} are a bad fit for exploratory programming.

\subsection{How External Sources Were Used}
We found the use of external sources to be very common. In Figure \ref{fig:codes}, we reported on the most frequent types of sources used. Participants frequently skimmed these sites and used them as external aids, and accessed them for short periods of time (browsing statistics per participant can be found on Table \ref{table:online_time}).

We also looked at what types of duplicates were most common in Figure \ref{fig:duplicates_taxonomy}. Given the results, it seems like a possible correlation exists between task familiarity and importance. Visualization snippets, for example, are frequently duplicated, because they are vital to the analysis process, but often unfamiliar to the analyst, who may not have extensive visualization training and instead relies on support from libraries such as Altair or GGPlot.

Duplication initially seems like a major time saver---since the chart, for instance, can be quickly reproduced---but eventually adds to technical debt and maintainability issues \cite{Thummalapenta:2009}. At that point, the duplication can be refactored into a common module, e.g., the corporate visualization module that defines fonts, themes, label sizes, etc., or common functions and classes used across notebooks. Commercial data science teams at places like Netflix have begun to support this process with extensive scaffolding around the basic notebook metaphor, for example, with Netflix's Metaflow\cite{Metaflow:0000} or AWS Step Functions\cite{AWS:0000}. Even further back, scientific workflow software \cite{Gil:2007} has been managing data processing models for many years.

\textbf{Practice implications}: Study 1 shows that it is difficult to understand when a code cell should be a module instead of duplicated code in a notebook. Software analytic techniques, applied in more conventional software programs, may help identify when duplicate code is becoming harmful \cite{Kapser:2008}.

\textbf{Research implications}: Social media and external (web) sources are widely recognized as a vital part of modern programming \cite{Storey:2014}. Programming support in notebooks should recognize this and support it, possibly by automating provenance and import, so that the original source can be referenced. In some samples, we saw this done manually with a comment referring to the Stack Overflow or API URL the solution was taken from.

\section{Limitations}
\noindent\textit{Construct Validity}. The main constructs we discuss are \textit{code duplicate} and \textit{code reuse}. We used the Levenshtein distance between code cells to detect duplicates, similar to Duala-Ekoko and Robillard in \cite{Duala_Ekoko:2007}, which is different than using an \textit{Abstract Syntax Tree (AST)}, which is more common in other code cloning research. This was due to the nature of Jupyter notebooks, which can support multiple kernels and programming languages, so we required a cross-language solution. We also set thresholds for duplicates, which are determined empirically based on soundness of the duplicates. However, for a different sample, these thresholds would provide sub-optimal results. In the future, an improvement would be to use a systematic analysis or grid search to find the parameters for duplicate detection and compare the detection results with an \textit{oracled} data set \cite{Bellon:2007}; this is especially true for $\lambda_1$ and $\lambda_2$, which control weights for the length and lines of code. However, we were not claiming general results for code duplication in all notebooks, and our findings should be seen as restricted to this sample. Also, given the emphasis we assigned on larger snippets and quantity of lines of code, there could be some under-reporting of duplicates, especially of shorter, more concise snippets of code, like short print statements, and other debugging techniques.

As we observed users to detect code reuse, it is possible that our small tasks did not truly test all forms of reuse. For example, our protocol did not allow for reuse from other participants. Similarly, our users were constrained to use our lab equipment. Using their own devices may have shown different reuse techniques (e.g., local copies of documentation).

\noindent\textit{Internal Validity}.
We used $897$ randomly selected repositories, consisting of $6,386$ notebooks and eight convenience sampled students. Our random sample of notebooks relied on the data set created by Rule in \cite{Rule:2018:2}. Because we sampled $897$ repositories and only considered inter-repository reuse, we may have missed reuse derived from external sources, such as popular repositories, tutorials, and training material. This almost certainly led to an under-reporting of code reuse.

Convenience sampling does not support statistical generalization, but since this was an exploratory study, generalization of the observed phenomena was not one of our objectives. While it is possible the behaviours we report on were unique to this sample, the participants all engaged in data analysis for several years in undergraduate and graduate courses at the university.

We asked for self-assessed proficiency with notebooks and version control. This has a potential observer-expectancy bias because we were known to the participants, who may have wanted to inflate their self-assessment. As an exploratory study, this is acceptable, but to test a specific hypothesis, our measures should be less prone to bias.

The tasks were scaled for the estimated skill of our participants. For example, \textit{np.mean(lst)} is sufficient to solve task 1 part 1, and for an experienced data scientist, could be retrieved from working memory. However, the tasks were designed for students, and if we were to use professionals, the tasks would have been more challenging. These tasks were designed to stimulate external/distributed cognition. We used three levels of complexity for our study, so that experienced analysts were given tasks commensurate with their skill. However, this is an imperfect matching process.

\noindent\textit{External Validity}.
We sampled from notebooks on GitHub, which may differ from notebooks used in corporate settings. Similarly, the use of students as subjects makes it difficult to draw generalizations about industry practitioners \cite{Feldt:2018}. That being said, the students in our sample reported using notebooks (and the other tools) frequently, and information on how professionals use notebooks is still limited.

%
%
\section{Conclusion}
We examined how code duplication and reuse happens in Jupyter notebooks. Our first study looked at how much self-duplication (i.e., within the repository) exists. We discovered that on average $7.6\%$ of code in repositories is self-duplicated. However, this did not explain how or from where code was duplicated to begin with.

We conducted a lab study with eight participants and deliberately crafted tasks designed to encourage reuse behaviour. We observed how participants reused code to solve data science tasks and how they leveraged version control, online sources and other notebooks. Reusing code from online sources proved to be the preferred method of reuse for our participants, with $18\%$ of their time spent browsing for code examples online, and version control systems proved to be the least effective method of reuse. Snippets of code that visualize data are the ones that are duplicated the most.

We conclude this paper by discussing observations and implications from our studies. First, while code duplication is clearly common in notebooks, the source of that duplication is important. Second, although much attention focuses on version control, for code reuse, other sources, such as API examples, are more important. Finally, these external sources are used for various tasks. Notebook interfaces should support modularization and reuse to improve cognitive support for data scientists.

\section{Acknowledgements}
Many thanks to Cassandra Petrachenko and Soroush Yousefi for their help with paper editing and data analysis, respectively. We also acknowledge the support of the Natural Sciences and Engineering Research Council of Canada (NSERC).

\bibliographystyle{IEEEtran}
\bibliography{Paper}

\end{document}